\documentclass[twocolumn,twoside,slac_two]{revtex4}
\usepackage{graphicx}
\usepackage{fancyhdr}
\begin{document}

\title{VERITAS Telescope 1 Relocation: Details and Improvements}

\author{Jeremy. S. Perkins}
\affiliation{Fred Lawrence Whipple Observatory, Harvard-Smithsonian
  Center for Astrophysics, Amado, AZ 85645, USA}
\author{Gernot Maier}
\affiliation{McGill University, 3600 University St., Montreal, QC H3A
  2TB, Canada}
\author{The VERITAS Collaboration}
\affiliation{For a list of authors see http://veritas.sao.arizona.edu}

\begin{abstract}
  The first VERITAS telescope was installed in 2002-2003 at the Fred
  Lawrence Whipple Observatory and was originally operated as a
  prototype instrument.  Subsequently the decision was made to locate
  the full array at the same site, resulting in an asymmetric array
  layout.  As anticipated, this resulted in less than optimal
  sensitivity due to the loss in effective area and the increase in
  background due to local muon initiated triggers.  In the summer of
  2009, the VERITAS collaboration relocated Telescope 1 to improve the
  overall array layout. This has provided a 30\% improvement in
  sensitivity corresponding to a 60\% change in the time needed to
  detect a source.
\end{abstract}

\maketitle

\thispagestyle{fancy}

\section{Introduction}

The imaging atmospheric Cherenkov technique (IACT) was developed at
the Fred Lawrence Whipple Observatory (FLWO) resulting in the first
very high energy (VHE; E $>$ 100 GeV) detection of the Crab nebula in
1989 \cite{Weekes:1989ph}.  In the twenty years since that first
publication there have been VHE detections of over 100 objects
including pulsars, blazars, pulsar wind nebula, supernova remnants and
starburst galaxies.  Since VHE photons do not penetrate the atmosphere
, IACT telescopes measure the Cherenkov light generated by particle
showers initiated by the primary photons interacting with our
atmosphere.  This Cherenkov light appears as a two dimensional ellipse
when imaged by an IACT telescope camera.  The shape and orientation of
the ellipse in the camera indicate whether the shower was initiated
by a gamma ray or by a cosmic ray which can also cause a particle shower.

The current generation of IACT instruments involve arrays of
telescopes.  The addition of multiple telescopes allows for a more
accurate determination of the shower parameters.  One of the most
powerful aspects of this technique is that the light pool of the
shower defines the collection area ($\sim 10,000~{\rm m}^2$) which is more
than adequate to compensate for the low flux of VHE gamma rays.
Currently there are four major experiments in operation, HESS, an
array of four IACT telescopes located in Namibia, MAGIC, an array of
two telescopes located in the Canary Islands, VERITAS in southern
Arizona and CANGAROO in Australia.  MAGIC just completed a major
upgrade by adding a single telescope and stereo trigger and HESS is in
the process of building an additional very large telescope.  This
contribution details part of the ongoing upgrade program being
undertaken by the VERITAS collaboration.

\begin{figure*}[t]
\centering
\includegraphics[width=135mm]{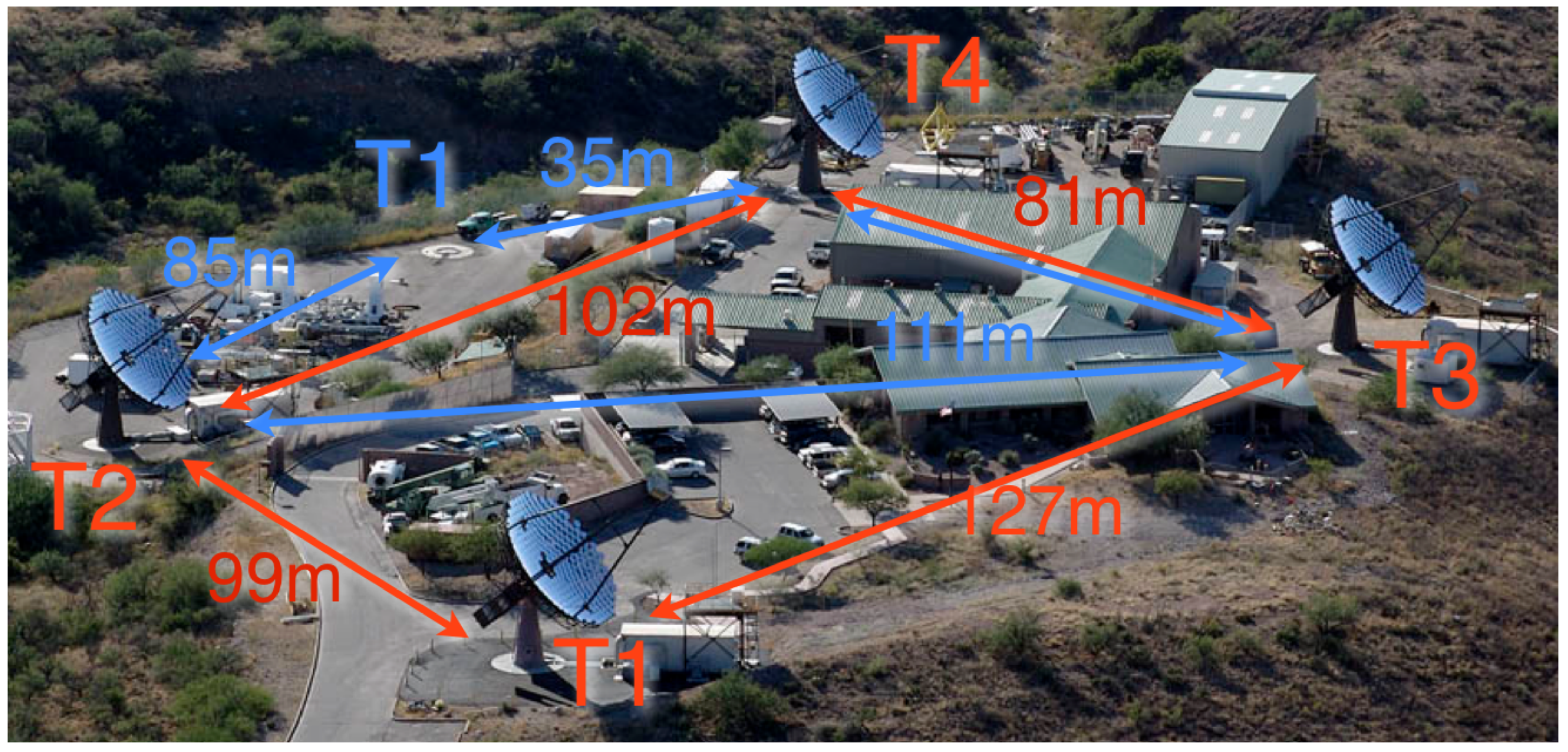}
\caption{aerial view of the new VERITAS array layout with Telescope 1
  relocated to the front of the FLWO administrative complex.  The
  original array layout position is marked in blue while the new one
  is in red.  Note the short distance (35m) between the original
  Telescope 1 position and the position of Telescope
  4.} \label{fig:layout}
\end{figure*}

VERITAS~\cite{Weekes:2002pi, Holder:2008rm} is an array of four 12 m
diameter IACT telescopes located in southern Arizona at the FLWO at an
altitude of 1268 m.  VERITAS detects photons from astrophysical
sources at energies between 100 GeV and 30 TeV.  The VERITAS
telescopes consist of four identical alt-az mounted Davies-Cotton
reflectors with an f number of 1.0.  The mirror area is approximately
106 m$^2$.  Mounted in the focal plane is a camera made up of 499
pixels consisting of 28 mm Photonis phototubes.  VERITAS has a three
level trigger, the first at the pixel level, the second is a pattern
trigger which triggers when any three adjacent pixels trigger.
Finally, an array trigger fires if any 2 or more telescopes trigger
within a set time frame.  For more details on the VERITAS hardware,
see \cite{Holder:2008rm}.

For historical reasons, Telescopes 1 and 4 were erected in close
($\sim$35 m) proximity.  Even though VERITAS met all of its original
design specifications, this resulted in a significant collection area
overlap and increased background due to cosmic rays and local muons.
In fact, all of the published VERITAS analysis included a cut that
rejected events that only triggered Telescopes 1 and 4.  Simulations
performed in the summer of 2008 suggested up to a 15\% improvement in
sensitivity if Telescope 1 was moved $\sim$200 m eastward from its
initial position.  Assuming that Telescopes 1 and 4 are redundant and
can be considered a single telescope, a 1/3 improvement is expected by
adding an additional telescope.

\begin{figure}
\includegraphics[width=65mm]{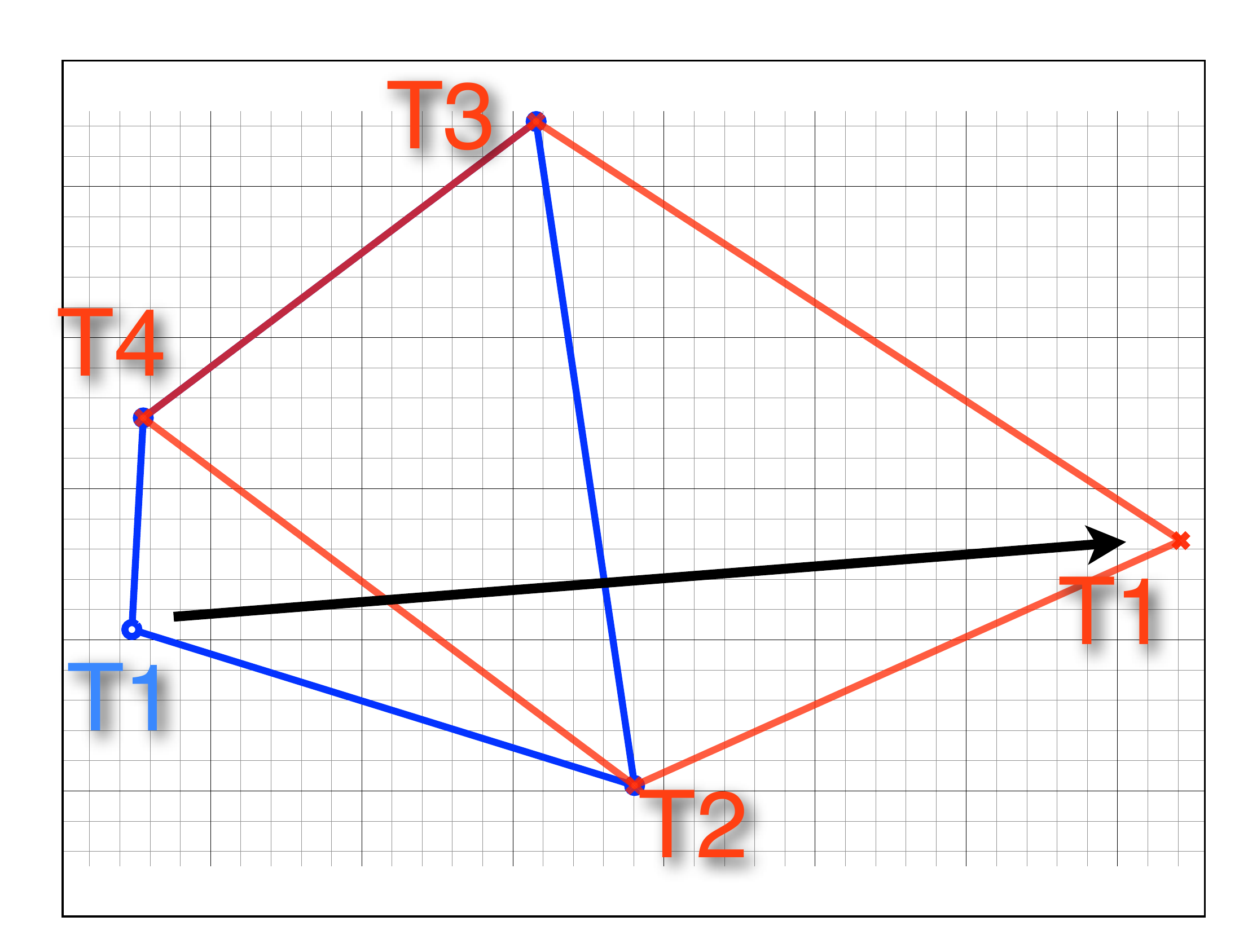}
\caption{Schematic of the array as viewed from directly overhead. The
  blue lines are for the original array layout while the red are for
  the new array layout. The black arrow indicates the relocation of
  Telescope 1. The small grid squares are 5 m on a side.}
\label{fig:layout-schematic}
\end{figure}

Based on these data, it was decided to relocate Telescope 1 to a more
ideal location providing a more symmetrical layout to the VERITAS
array (see Figures \ref{fig:layout} and \ref{fig:layout-schematic}).
It was decided to relocate Telescope 1 instead of Telescope 4 to allow
for the refurbishment of the oldest telescope in the array which was
originally installed at the FLWO as a prototype in 2002. The
relocation of Telescope 1 is part of an ongoing upgrade program
\cite{Nepomuk-Otte:2009qy} which recently included an improvement in
the optical point spread function (PSF) \cite{McCann:2009fk}.  The
improvement in the optical PSF was accomplished using a novel mirror
alignment system which resulted in a 25 - 30\% improvement in the PSF.
This optical PSF improvement also contributes to the enhancement in
sensitivity discussed here and cannot be disentangled from the overall
results. The move of Telescope 1 combined with the improvement in the
optical PSF has resulted in making VERITAS the most sensitive VHE
telescope array in the world capable of detected a 1\% crab nebula
signal in less than 30 hours.

Since VERITAS does not operate during the summer months (approximately
July through August), the move of Telescope 1 was scheduled to take
place during this time to minimize the impact on the observing
program.  Telescope 1 was shutdown 6 weeks early (May 4, 2009) so that
it would be operational by the first of October.  The move was
completed on September 4, 2009 and is estimated to have taken 2600
person hours of labor.  Ten days later on the $14^{\rm th}$ scheduled
operations began with the full array, over two weeks earlier than
expected.  By September $17^{\rm th}$ normal operations had resumed.
In total, VERITAS only lost 6 weeks of full four telescope operations
and these were with the old array layout.  The final array layout,
while not entirely symmetric, is a much better layout for a VHE
instrument.  Figure \ref{fig:layout} shows an aerial view of the
VERITAS array with the old layout shown in blue and the new layout in
red.  While the old layout had inter-telescope distances ranging from
35 m to 127 m, the new layout distances range from 81 m to 127 m.
Figure \ref{fig:layout-schematic} shows a schematic representation of
the array viewed from directly above.  Also shown as a black arrow is
the relocation of Telescope 1.

\section{Results}

\begin{figure}
\includegraphics[width=65mm]{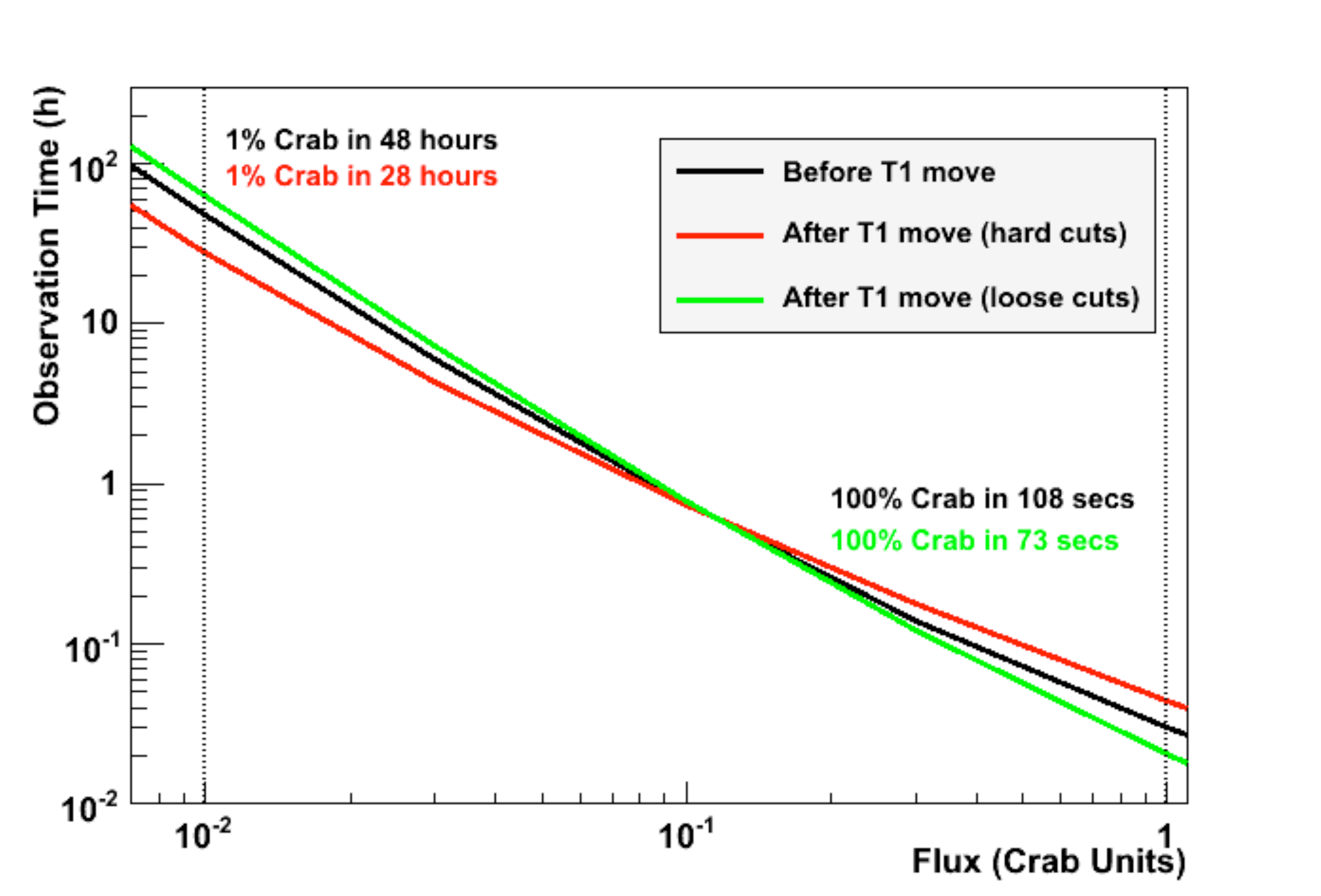}
\caption{The time needed to detected a source at the 5 $\sigma$ level
  vs. that source's flux in units of the Crab Nebula's flux.  This is
  shown for the original array layout and for the new array layout
  with two different sets of event selection cuts.  Note that it would
  take 48 hours to detect a 1\% Crab source with the original array
  layout and only 28 hours with the new array layout.}
\label{fig:time}
\end{figure}

VERITAS data are calibrated and cleaned initially as described in
\cite{Daniel:2008lr}.  After calibration several noise-reducing cuts
are made.  The VERITAS standard analysis consists of parametrization
using a moment analysis \cite{Hillas:1985ta} and following this, the
calculation of scaled parameters are used for event selection
\cite{Aharonian:1997rm,Krawczynski:2006ts}. This selection consists
of different sets of gamma-ray cuts, determined {\it a priori}.
Depending on the strength and expected spectral index of a source,
different cuts are chosen.  For example, a source with the strength of
the Crab would use a looser set of cuts than a week source at the 1\%
Crab flux level.  These two sets of cuts are called loose and hard
cuts.  Additionally, soft and standard versions of these cuts are used
for soft (approximately spectral indices of 3 and above) or standard
(Crab-like 2.5 spectral index sources).  The choice of which cuts to
use are determined prior to the analysis.  Figure \ref{fig:time} shows
the observation time needed to detect an object at the 5 standard
deviation ($\sigma$) level.  Before the relocation of Telescope 1, a
1\% Crab flux source could be detected in 48 hours while after the
move it only takes 28 hours loose cuts.  Similarly, it takes 72
seconds to detect the Crab nebula after the move with hard cuts, as
opposed to 108 seconds before the move.  

\begin{figure}
\includegraphics[width=65mm]{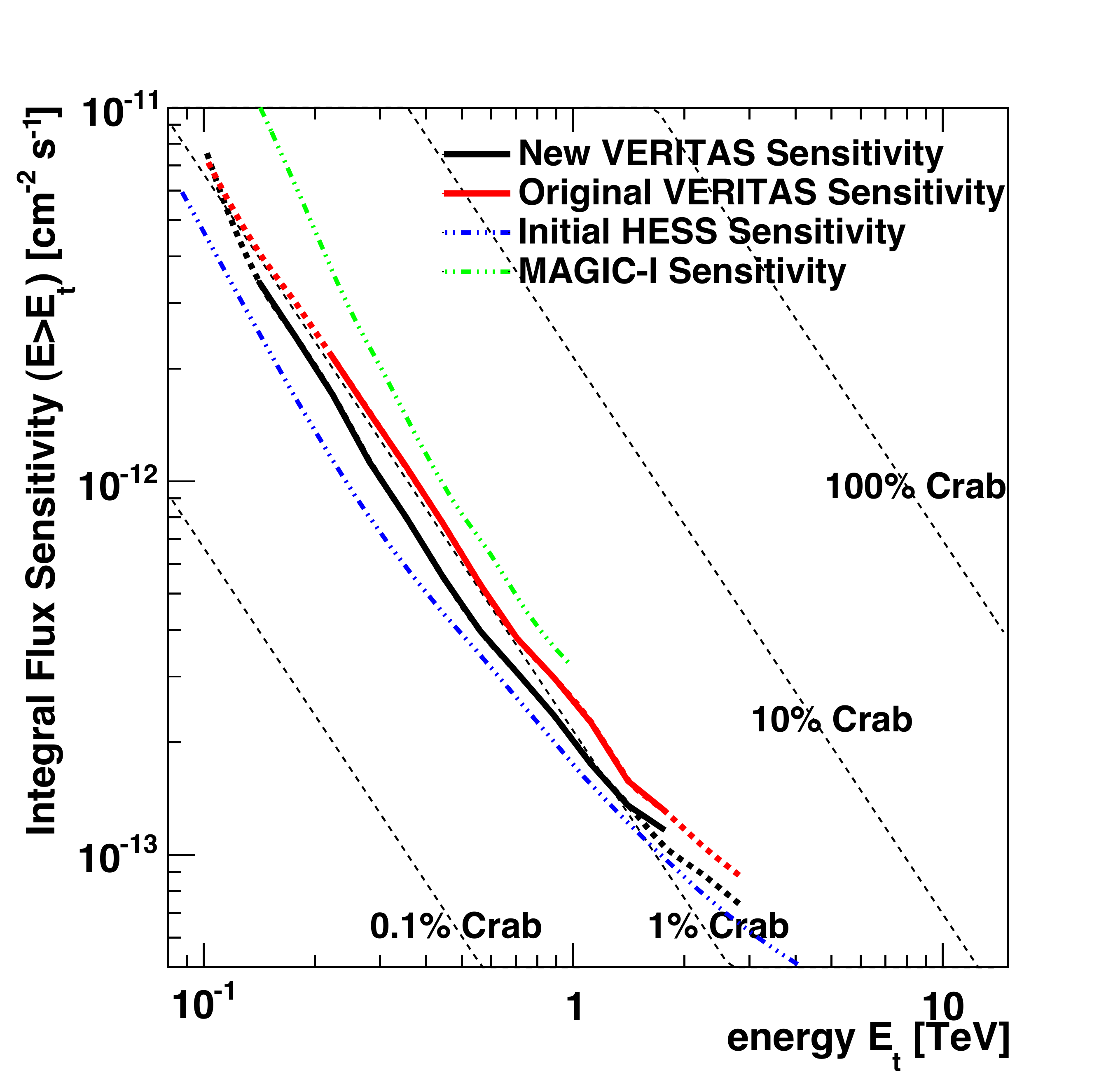}
\caption{Integral flux sensitivity vs. Energy Threshold for several
  different instruments.  In red is the original VERITAS layout while
  the new VERITAS layout is shown in black (based on Crab
  observations; the dashed sections are under evaluation).  Initial
  HESS sensitivity \cite{Aharonian:2006kk} is shown as the blue dashed
  line and MAGIC-I$^1$ is shown as the green dashed line.  Note that
  the integral flux sensitivity improvement of VERITAS above 300 GeV
  is $\sim30\%$.}
\label{fig:sensitivity}
\end{figure}

Another way of looking at the sensitivity of VHE instruments is to
calculate the integral flux sensitivity above a energy threshold.
Shown in Figure \ref{fig:sensitivity} are the integral flux
sensitivity vs. Energy Threshold for several different instruments.
In red is the original VERITAS layout while the new VERITAS layout is
shown in black (based on Crab observations; the dashed sections are
under evaluation). The integral flux sensitivity of VERITAS above 300
GeV is $\sim30\%$ better after the move.  This corresponds directly to
a 60\% reduction in the time needed to detect a source (for example, a
50 hour observation before the move is equivalent to a 30 hour post
move observation). For comparison are shown the initial HESS
sensitivity \cite{Aharonian:2006kk} shown as the blue dashed line
(note that this is the original HESS sensitivity curve before any
mirror reflectivity degradations).  The integral sensitivity of a
single MAGIC-I\footnote{http://www.astro.uni-wuerzburg.de/mphysics/}
telescope is shown as the green dashed line.  Another thing to note is
that the sensitivity of VERITAS has slightly degraded at the lower end
due to the loss of sensitivity to the lowest energy showers.

\begin{figure*}[t]
\centering
\includegraphics[width=135mm]{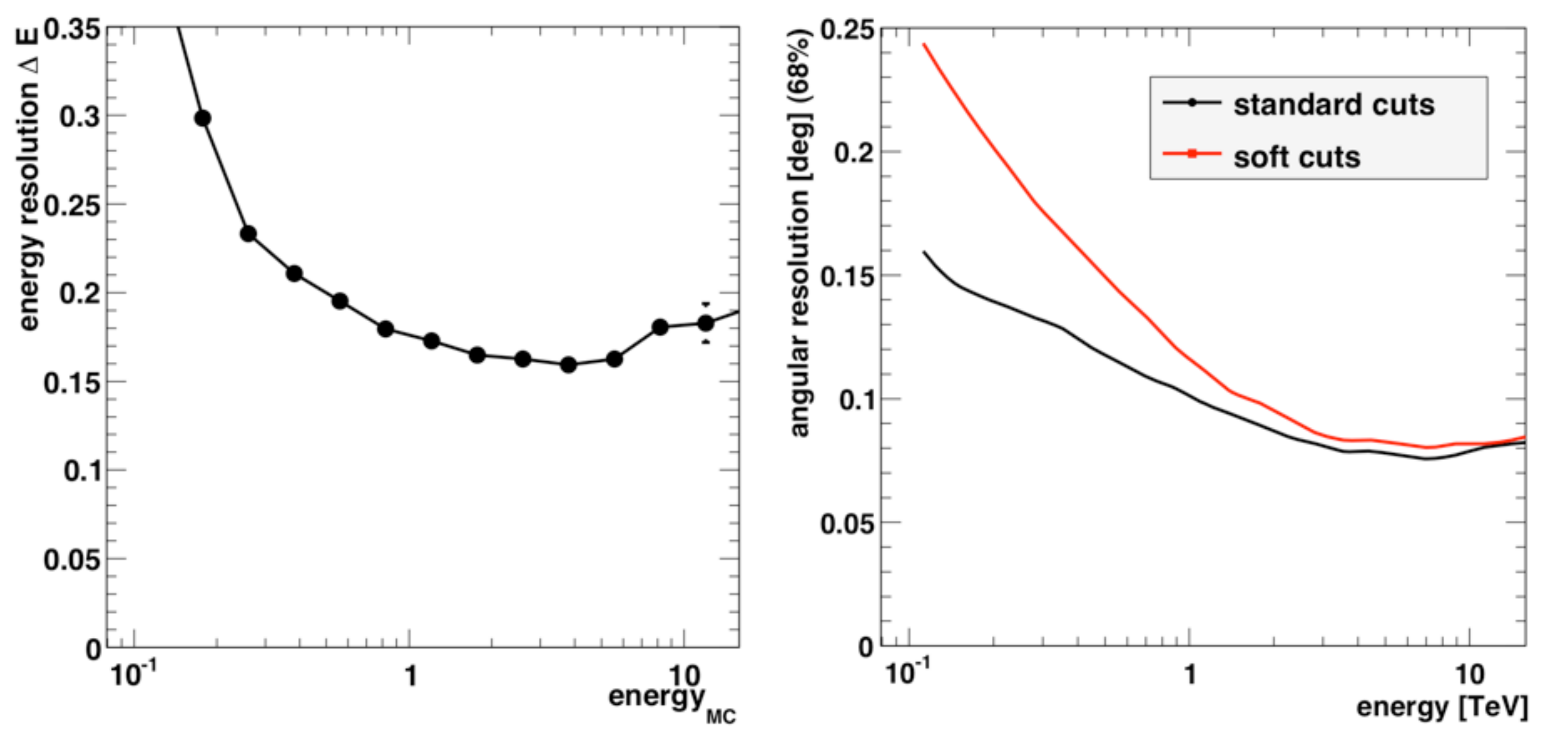}
\caption{Left: the energy resolution of VERITAS vs. simulated energy
  at 70 degrees.  Right: the angular resolution (68\% containment) of
  VERITAS vs. real energy at an elevation of 70 degrees for two
  different sets of event selection cuts (black is for a crab like
  source and red is for a softer spectrum source).  }
\label{fig:resolution}
\end{figure*}

Figure \ref{fig:resolution} shows the energy resolution of the VERITAS
array as well as the angular resolution.  These two features are
similar to the numbers calculated for the original VERITAS array
layout.  Both plots are for observations at 70 degrees elevation.  The
angular resolution and energy resolution change for observations
undertaken at different elevations.

\section{Summary}
The VERITAS collaboration relocated Telescope 1 and dramatically
improved the optical PSF during the summer of 2009 as part of an
ongoing upgrade program.  These studies indicate that the upgrades
have improved the sensitivity of VERITAS by 30\% resulting in a 60\%
change in the time needed to detect a source.  The higher sensitivity
achieved with VERITAS allows the detection of more objects in a
shorter amount of time, effectively doubling the observation time.
The ability to detect marginal sources such as M82 and to do deep
observations of known objects has drastically improved.

In addition to the telescope relocation and optical PSF improvements,
there are several other upgrade plans being discussed which are
described in \citep{Nepomuk-Otte:2009qy} and are planned to be
implemented in the next few years.  These upgrade plans include the
installation of higher efficiency photon detectors which would result
in a 17\% improvement in the sensitivty of the array and/or the
installation of a topological trigger which would consist of
transmitting image parameters from the camera directly to the array
trigger allowing for real-time event classification for gamma/hadron
separation.  In addition to these baseline upgrades, the expansion of
the array by adding more telescopes or an active mirror alignment
system is also possible.  This ongoing upgrade program, beginning with
the optical PSF improvement and the relocation of Telescope 1 will
continue to make VERITAS competitive in the coming decade.

\bigskip

\begin{acknowledgments}
  This research is supported by grants from the U.S. Department of
  Energy, the U.S. National Science Foundation, and the Smithsonian
  Institution, by NSERC in Canada, by PPARC in the UK and by Science
  Foundation Ireland.  The VERITAS collaboration acknowledges the hard
  work and dedication of the FLWO support staff in making the
  relocation of Telescope 1 possible.
\end{acknowledgments}

\bigskip


\end{document}